\documentclass[a4paper,oneside,final,notitlepage,onecolumn]{article}
\usepackage{amsfonts}

\begin{document}

\newtheorem{theorem}{Theorem}[section]
\newtheorem{lemma}{Lemma}[section]
\newtheorem{proposition}{Proposition}[section]
\newtheorem{corollary}{Corollary}[section]
\newtheorem{conjecture}{Conjecture}[section]
\newtheorem{example}{Example}[section]
\newtheorem{definition}{Definition}[section]
\newtheorem{remark}{Remark}[section]
\newtheorem{exercise}{Exercise}[section]
\newtheorem{axiom}{Axiom}[section]
\renewcommand{\theequation}{\thesection.\arabic{equation}} 

\author{Helmut Friedrich\thanks{%
~email: hef@aei-potsdam.mpg.de} \\
Max-Planck-Institut f\"{u}r Gravitationsphysik, Albert-Einstein-Institute\\
Schlaatzweg 1, 14473 Potsdam, Germany 
\and Istv\'{a}n R\'{a}cz\thanks{%
~Fellow of the Japan Society for the Promotion of Science, on leave of absence
from MTA-KFKI Research Institute for Particle and Nuclear Physics, email: 
istvan@yukawa.kyoto-u.ac.jp} 
\\ 
Yukawa Institute for Theoretical Physics\\
Kyoto University, Kyoto 606-01, Japan
\and Robert M. Wald\thanks{%
~email: rmwa@midway.uchicago.edu} \\
Enrico Fermi Institute, University of Chicago, 5640 S. Ellis Ave.\\
Chicago, IL 60637-1433, USA} 
\title{On the Rigidity Theorem for Spacetimes with a Stationary Event
Horizon or a Compact Cauchy Horizon} 
\maketitle

\begin{abstract}

We consider smooth electrovac spacetimes which represent either (A) an
asymptotically flat, stationary black hole or (B) a cosmological
spacetime with a compact Cauchy horizon ruled by closed null
geodesics. The black hole event horizon or, respectively, the compact
Cauchy horizon of these spacetimes is assumed to be a smooth null
hypersurface which is non-degenerate in the sense that its null
geodesic generators are geodesically incomplete in one direction. In
both cases, it is shown that there exists a Killing vector field in a
one-sided neighborhood of the horizon which is normal to the
horizon. We thereby generalize theorems of Hawking (for case (A)) and
Isenberg and Moncrief (for case (B)) to the non-analytic case.

\end{abstract}

\section{Introduction}
\setcounter{equation}{0}

A key result in the theory of black holes is a theorem of Hawking
\cite{hawk1, HE} (see also \cite{chrus22,chrus3}), which asserts that,
under certain hypotheses, the event horizon of a stationary,
electrovac black hole is necessarily a Killing horizon, i.e., the
spacetime must possess a Killing field (possibly distinct from the
stationary Killing field) which is normal to the event horizon.  The
validity of this theorem is of crucial importance in the
classification of stationary black holes, since it reduces the problem
to the cases covered by the well-known uniqueness theorems for
electrovac black holes in general relativity
\cite{israel1,israel2,carter1,carter2,mazur,bunting}. However, an important,
restrictive hypothesis in Hawking's theorem is that the spacetime be
analytic.

A seemingly unrelated theorem of Isenberg and Moncrief \cite{im1,im2}
establishes that in an electrovac spacetime possessing a compact
Cauchy horizon ruled by closed null geodesics, there must exist a
Killing vector field which is normal to the the Cauchy horizon. This
result supports the validity of the strong cosmic censorship hypothesis
\cite{P2} by demonstrating that the presence of such a compact Cauchy
horizon is ``non-generic''. The Isenberg-Moncrief theorem also
contains the important, restrictive hypothesis that the spacetime be
analytic.

The main purpose of this paper is to show that the theorems of Hawking
and of Isenberg and Moncrief can be proven in the case of a smooth (as
opposed to analytic) geometrical setting.\footnote{Further
generalizations to allow for the presence of other types of matter
fields will be treated by one of us elsewhere \cite{racz}.}
However, a fundamental limitation of our method is that we are
able to prove existence of the Killing field only on a one-sided
neighborhood of the relevant horizon. For the Hawking theorem, this
one-sided neighborhood corresponds to the interior of the black hole,
whereas the existence of a Killing field in the exterior region is
what is relevant for the black hole uniqueness theorems. However, for
the Isenberg-Moncrief theorem, the one-sided neighborhood corresponds
to the original Cauchy development, so our results significantly
strengthen their conclusion that the presence of a compact Cauchy
horizon ruled by closed null geodesics is an artifact of a spacetime
symmetry.

This paper is organized as follows: In Section 2 we consider
stationary black hole spacetimes and establish the existence of a
suitable discrete isometry which maps each generator of the event
horizon into itself. As seen in Section 3, by factoring the spacetime
by this isometry, we produce a spacetime having the local geometrical
properties of the spacetimes considered by Isenberg and Moncrief. This
construction explicitly demonstrates the close mathematical
relationship between the Hawking and Isenberg-Moncrief theorems. In
Section 3, we also review the relevant result of Isenberg and
Moncrief, which shows that in suitably chosen Gaussian null coordinates
defined in the ``unwrapping'' of certain local neighborhoods covering
the horizon, $\mathcal{N}$, all the fields and their
coordinate derivatives transverse to $\mathcal{N}$ are independent of
the coordinate $u$ on $\mathcal{N}$. In the analytic case, this
establishes that $(\partial/\partial u)^a$ is (locally) a Killing
field. Section 4 contains the key new idea of the paper: We use the
methods of \cite{rw1}, \cite{rw2} to extend the region covered by the
local Gaussian null coordinates of Isenberg and Moncrief so that the
extended spacetime is smooth and possesses a bifurcate null
surface. This bifurcate null surface then provides a suitable initial
data surface, from which the existence of a Killing field on the
extended (and, hence, on the original) spacetime can be established
without appealing to analyticity. The results concerning the null
initial value formulation that are needed to establish the existence
of a Killing field are proven in an appendix.

Throughout this paper a spacetime $(M,g_{ab})$ is taken to be a
smooth, paracompact, connected, orientable manifold $M$ endowed with a
smooth Lorentzian metric $g_{ab}$ of signature $(+,-,-,-)$. It is
assumed that $(M,g_{ab})$ is time orientable and that a time
orientation has been chosen. The Latin indices $a,b,c...$ will be
used as abstract tensor indices \cite{wald}, the Latin indices
$i,j,k...$ will denote tetrad components (used only in
Appendix B), and Greek indices will denote coordinate components.

\section{Stationary black hole spacetimes}
\setcounter{equation}{0}

In this section we shall give a mathematically precise specification
of the class of stationary, black hole spacetimes to be considered,
and we then shall prove existence of a discrete isometry which maps
each generator of the horizon into itself.

We consider smooth, strongly causal spacetimes $(M,g_{ab})$ which are
$(k,\alpha)$-asymptotically stationary as specified in definition 2.1
of \cite{CW2}. Thus, we assume that $(M,g_{ab})$ possesses a
one-parameter group of isometries, $\phi_t$, generated by a Killing
vector field $t^a$, and possesses a smooth acausal slice $\Sigma$
which contains an asymptotically flat ``end'', $\Sigma_{end}$, on
which $t^a$ is timelike and the properties specified in definition 2.1
of \cite{CW2} hold. However, the precise asymptotic flatness
conditions given in that definition will not be of great importance
here, and could be significantly weakened or modified. We further
require that if any matter fields (such as an electromagnetic field,
$F_{ab}$) are present in the spacetime, then they also are invariant
under the action of $\phi_t$.

We define $M_{end}$ to be the orbit\footnote{%
The orbit of an arbitrary subset $Q\subset M$ under the action of $\phi_t$ 
is defined to be $\phi\{Q\}=\cup_{t\in \mathbb R}\phi_t[Q]$.}
of $\Sigma_{end}$ under the isometries 
\begin{equation}
M_{end}=\phi\{\Sigma_{end}\}.
\end{equation}
The black hole region $\mathcal{B}$ is defined to be the complement of
$I^-[M_{end}]$ and the white hole region $\mathcal{W}$ is defined to be 
the complement
of $I^+[M_{end}]$. We require that $(M,g_{ab})$ possess a black hole
but no white hole, i.e. $\mathcal{W}=\emptyset$ which implies that
\begin{equation}
M=I^+[M_{end}].
\label{nowhite}
\end{equation}
Note that the domain of outer communications
$\mathcal{D}$ associated with the asymptotically flat end is, in
general, defined to be the intersection of the chronological future
and past of $M_{end}$, but, in view of eq.(\ref{nowhite}), we have
simply
\begin{equation}
\mathcal{D}=I^-[M_{end}].
\end{equation}

The (future) event horizon of the spacetime is defined by 
\begin{equation}
\mathcal{N}=\partial I^-[M_{end}].
\end{equation}
Our final requirement is that $\mathcal{N}$ is smooth and that the manifold
of null geodesic generators of $\mathcal{N}$ has topology $S^2$ (so
that $\mathcal{N}$ has topology $\mathbb{R} \times S^2 $).

\begin{definition}
Stationary black hole spacetimes which satisfy all of the above
assumptions will be referred as spacetimes of {\it class A}.
\end{definition}

\begin{remark}
If it is merely assumed that $\mathcal{N}$ has topology $\mathbb{R}
\times K $, where $K$ is compact, then under some mild additional
assumptions, it follows from the topological censorship theorem
\cite{FSW} that each connected component of $\mathcal{N}$ has topology
$\mathbb{R} \times S^2 $ \cite{CW,G1,G2}; see remark \ref{CWproof}
below for a strengthening of this result. However, rather than
introduce any additional assumptions here, we have chosen to merely
assume that $\mathcal{N}$ has topology $\mathbb{R} \times S^2$.
\end{remark}

We begin with the following lemma

\begin{lemma}\label{fixedpt}
Let $(M,g_{ab})$ be a spacetime of class A. Then for all $q \in
\mathcal{N}$ and all $t \neq 0$ we have $\phi_t (q) \neq q$. In
particular, $t^a$ is everywhere non-vanishing on $\mathcal{N}$.
\end{lemma}

\noindent
{\bf Proof}{\ } Suppose that for some $q \in \mathcal{N}$ and some $t
\neq 0$ we had $\phi_t (q) = q$. Since $M=I^+[M_{end}]$, there exists
$p \in M_{end}$ such that $p \in I^-(q)$.  Since $\phi_{nt}(q) = q$
for all integers $n$, it follows that $\phi_{nt} (p) \in I^-(q)$ for
all $n$, from which it follows that $\phi \{p\} \in
I^-(q)$. Therefore, by lemma 3.1 of \cite{CW2}, we have $I^-(q)
\supset M_{end}$ and, hence, $I^-(q) \supset I^-[M_{end}] =
\mathcal{D}$. However, since $\mathcal{N}=\partial I^-[M_{end}]$, it
follows that $q$ lies on a future inextendible null geodesic,
$\gamma$, contained within $\mathcal{N}$. Let $r$ lie to the future of
$q$ along $\gamma$. Let $O$ be an open neighborhood of $r$ which does
not contain $q$ and let $V$ be any open neighborhood of $r$ with $V
\subset O$. Since $r \in \mathcal{N}=\partial I^-[M_{end}] = \partial
\mathcal{D}$, we have $V \cap \mathcal{D} \neq \emptyset$. Hence we
can find a causal curve which starts in $V \cap \mathcal{D}$, goes to
$q$ (since $I^-(q) \supset \mathcal{D}$) and then returns to $r$ along
$\gamma$. This violates strong causality at $r$. \hfill \fbox{}

\begin{remark} \label{CWproof}
No assumptions about the topology or smoothness of $\mathcal{N}$ were
used in the proof of this lemma. It is worth noting that a step in the
proof of this lemma can be used to strengthen the results of
\cite{CW}, so as to eliminate the need for assuming existence of an
asymptotically flat slice that intersects the null geodesic generators
of $\mathcal{N}$ in a cross section. First, we note that part (1) of
lemma 2 of \cite{CW} can be strengthened to conclude that for any $p
\in M_{end}$, each Killing orbit, $\alpha$, on $\mathcal{N}$
intersects the achronal $C^{1-}$ hypersurface $\mathcal{C} \equiv
\partial I^+(p)$ in precisely one point. (Lemma 2 of \cite{CW} proved
the analogous result for Killing orbits in $\mathcal{D}$.) Namely by
Lemma 3.1 of \cite{CW2}, $\alpha$ satisfies either $I^-[\alpha]\cap
M_{end}=\emptyset$ or $I^-[\alpha]\supset M_{end}$. The first
possibility is excluded by our assumption that $M=I^+[ M_{end}]$, so
there exists $q\in \alpha \cap I^+(p)$. On the other hand, the proof
of the above lemma shows that $I^-(q)$ cannot contain $\mathcal{D}$,
so there exists $t > 0$ such that $q \not \in I^+(\phi_t
(p))$. Equivalently, we have $\phi_{-t} (q) \not \in I^+(p)$, which
implies that the Killing orbit $\alpha$ must intersect
$\mathcal{C}$. Furthermore, if $\alpha$ intersected $\mathcal{C}$ in
more than one point there would exist $t>0$ so that both $r \in
\alpha$, and $\phi_t(r)$ lie on $\mathcal{C}$.  This would imply, in
turn, that $r$ lies on the boundary of both $I^+(p)$ and
$I^+(\phi_{-t}(p))$ which is impossible since $p\in
I^+(\phi_{-t}(p))$. Consequently, each Killing orbit on $\mathcal{N}$
intersects $\mathcal{C}$ precisely once, i.e., $\varsigma \equiv
\mathcal{C} \cap \mathcal{N}$ is a cross-section for the Killing
orbits, as we desired to show.\footnote{Furthermore, if $\mathcal{N}$
is smooth and $\varsigma$ is compact, then $\varsigma$ also is a
cross-section for the null geodesic generators of
$\mathcal{N}$. Namely, smoothness of $\mathcal{N}$ precludes the
possibility that a null geodesic generator, $\gamma$, of
$\mathcal{N}$ has endpoints. (Future endpoints are excluded in any
case, since $\mathcal{N}$ is a past boundary.) To show that
$\gamma$ must intersect $\varsigma$, let $r \in \gamma$ and let $t$ be
such that $\phi_{-t} (r) \in \varsigma$. (Such a $t$ exists since
$\varsigma$ is a cross section for Killing orbits.) Suppose that $t >
0$. If $\gamma$ failed to intersect $\varsigma$, then the segment of
$\gamma$ to the past of $r$ would be a past inextendible null geodesic
which is confined to the compact region bounded by $\varsigma$ and
$\phi_t [\varsigma]$. This violates strong causality. Similar arguments
apply for the case where $t < 0$, thus establishing that $\gamma$ must
intersect $\varsigma$. Finally, if $\gamma$ intersected $\varsigma$ at
two points, $q,s$, then by achronality of $\mathcal{C}$, the segment
of $\gamma$ between $q$ and $s$ must coincide with a null geodesic
generator, $\lambda$, of $\mathcal{C}$. When extended maximally into
the past, this geodesic must remain in $\mathcal{N}$ (by smoothness of
$\mathcal{N}$) and in $\mathcal{C}$ (since $\mathcal{C}$ is a future
boundary and $p \not \in \mathcal{N}$). Thus, we obtain a past
inextendible null geodesic which lies in the compact set $\varsigma =
\mathcal{N} \cap \mathcal{C}$, in violation of strong causality.} In
particular, this shows that $\mathcal{N}$ has the topology $\mathbb{R}
\times \varsigma$. If we now assume, as in \cite{CW}, that
$\mathcal{D}$ is globally hyperbolic, that the null energy condition
holds, and that $[\mathcal{C} \setminus \mathcal{C}_{ext}] \cap
\mathcal{D}$ has compact closure in $M$, then the same argument as
used in the proof of Theorem 3 of \cite{CW} establishes that each
connected component of $\varsigma$ has topology $S^2$, without the
need to assume the existence of an achronal slice which intersects the
null geodesic generators of $\mathcal{N}$ in a cross section.
\end{remark}

Our main result of this section is the following.\footnote{Note that
the conclusion of this Proposition was assumed to be satisfied in the
proof of Prop. 9.3.6 of \cite{HE}, but no justification for it was provided
there.}

\begin{proposition}\label{prop-orbits}
Let $(M,g_{ab})$ be a spacetime of class A which satisfies the null
energy condition, $R_{ab} k^a k^b \geq 0$ for all null $k^a$. Then
there exists a $t_0 \neq 0$ such that $\phi_{t_0}$ maps each null
geodesic generator of $\mathcal{N}$ into itself. Thus, the Killing
orbits on $\mathcal{N}$ repeatedly intersect the same generators with
period $t_0$.
\end{proposition}

\noindent{\bf Proof}{\ } By Proposition 9.3.1 of \cite{HE}, the
expansion and shear of the null geodesic generators of $\mathcal{N}$
must vanish. By Lemma \ref{lem-rs=0} of Appendix B, this implies that ${\cal
L}_k\,{g'}_{ab} = 0$ on $\mathcal{N}$, where ${g'}_{ab}$ denotes the
pullback of $g_{ab}$ to $\mathcal{N}$ and $k^a$ is any smooth vector
field normal to $\mathcal{N}$ (i.e., tangent to the null geodesic
generators of $\mathcal{N}$). Since we also have ${g'}_{ab} k^b = 0$,
it follows from the Appendix of \cite{geroch} that ${g'}_{ab}$ gives
rise to a negative definite metric, $\hat{g}_{AB}$, on the manifold,
$\mathcal{S}$, of null geodesic orbits of $\mathcal{N}$. By our
assumptions, $\mathcal{S}$ has topology $S^2$.

Now, for all $t$, $\phi_t$ maps $\mathcal{N}$ into itself and also
maps null geodesics into null geodesics. Consequently, $\phi_t$ gives
rise to a one parameter group of diffeomorphisms $\hat{\phi}_t$ on
$\mathcal{S}$, which are easily seen to be isometries of
$\hat{g}_{AB}$. Let $\hat{t}^A$ denote the corresponding Killing field
on $\mathcal{S}$. If $\hat{t}^A$ vanishes identically on $\mathcal{S}$
(corresponding to the case where $t^a$ is normal to $\mathcal{N}$),
then the conclusion of the Proposition holds for all $t_0 \neq 0$. On
the other hand, if $\hat{t}^A$ does not vanish identically, then since
the Euler characteristic of $\mathcal{S}$ is non-vanishing, there
exists a $p \in \mathcal{S}$ such that $\hat{t}^A (p) = 0$.  By the
argument given on pages 119-120 of \cite{W2}, it follows that there exists
a $t_0 \neq 0$ such that $\hat{\phi}_{t_0}$ is the identity map on
$\mathcal{S}$. Consequently, $\phi_{t_0}$ maps each null geodesic
generator of $\mathcal{N}$ into itself. \hfill\fbox{}

\section{Isenberg-Moncrief Spacetimes}
\setcounter{equation}{0}

In this Section, we shall consider spacetimes, $(M,g_{ab})$, which
contain a compact, orientable, smooth null hypersurface,
$\mathcal{N}$, which is generated by closed null geodesics.

\begin{definition}
Spacetimes which satisfy the above properties will be referred to as
spacetimes of {\it class B}.
\end{definition}

Spacetimes of class B arise in the cosmological context. In
particular, the Taub-NUT spacetime and its generalizations given in
Refs. \cite {mill,monc1,monc2}) provide examples of spacetimes of
class B. In these examples, $\mathcal{N}$ is a Cauchy horizon which
separates a globally hyperbolic region from a region which contains
closed timelike curves. However, it should be noted that even among
the Kerr-Taub-NUT spacetimes one can find (see Ref. \cite{mill})
spacetimes with compact Cauchy horizons for which almost all of the
generators of the horizon are not closed. Therefore it should be
emphasized that here we restrict consideration to horizons
foliated by circles.

Since strong causality is violated in all spacetimes of class B, it is
obvious that no spacetime of class A can be a spacetime of class
B. Nevertheless, the following Proposition shows that there is a very
close relationship between spacetimes of class A and spacetimes of
class B:

\begin{proposition}\label{AB}
Let $(M,g_{ab})$ be a spacetime of class A. Then there exists an open
neighborhood, $\mathcal{O}$, of the horizon, $\mathcal{N}$, such that
$(\mathcal{O}, g_{ab})$ is a covering space of a spacetime of class B.
\end{proposition}

\noindent{\bf Proof}{\ } Let $t_0 > 0$ be as in Proposition
\ref{prop-orbits}. By Lemma \ref{fixedpt}, $\phi_{t_0}$ has no fixed
points on $\mathcal{N}$. Since the fixed points of an isometry
comprise a closed set, there exists an open neighborhood, ${\cal U}$,
of $\mathcal{N}$ which contains no fixed points of $\phi_{t_0}$. Let
$\mathcal{O} = \phi\{{\cal U}\}$. Then clearly $\mathcal{O}$ also is an open
neighborhood of $\mathcal{N}$ which contains no fixed points of
$\phi_{t_0}$. Moreover, $\phi_{t_0}$ maps $\mathcal{O}$ into itself. Let
$(\tilde{M}, \tilde{g}_{ab})$ be the factor space of $(\mathcal{O},
g_{ab})$ under the action of the isometry $\phi_{t_0}$. Then
$(\tilde{M}, \tilde{g}_{ab})$ is a spacetime of class B, with covering
space $(\mathcal{O},g_{ab})$.  \hfill\fbox{}

\medskip

Now, if a spacetime possesses a Killing vector field $\xi^a$, then any
covering space of that spacetime possesses a corresponding Killing
${\xi'}^a$ that projects to $\xi^a$. Consequently, if the existence of a
Killing field is established for spacetimes of class B, it follows
immediately from Proposition \ref{AB} that a corresponding Killing
field exists for all spacetimes of class A. In particular, for
analytic electrovac spacetimes of class B, Isenberg and Moncrief
\cite{im1,im2} proved existence of a Killing field in a neighborhood
of $\mathcal{N}$ which is normal to $\mathcal{N}$. Consequently, for
any analytic, electrovac spacetime of class A, there also exists a
Killing field in a neighborhood of $\mathcal{N}$ which is normal to
$\mathcal{N}$. Thus, Hawking's theorem \cite{hawk1, HE} may be
obtained as a corollary of the theorem of Isenberg and Moncrief
together with Proposition \ref{AB}.

The main aim of our paper is to extend the theorems of Hawking and of
Isenberg and Moncrief to the smooth case. In view of the above remark,
it suffices to extend the Isenberg-Moncrief theorem, since the
extension of the Hawking theorem will then follow automatically. Thus,
in the following, we shall restrict attention to spacetimes of
class B.

For spacetimes of class B, $\mathcal{N}$ is a compact, orientable
3-manifold foliated by closed null geodesics.  To discuss this
situation we introduce some terminology (cf. \cite{se} for more
details). The ``ordinary fibered solid torus'' is defined as the set
$D^2 \times S^1$ with the circles $\{p\} \times S^1$, $p \in D^2$, as
``fibers''. Here $D^2$ denotes the 2-dimensional closed unit disk. A
``fibered solid torus'' is obtained by cutting an ordinary fibered
solid torus along a disk $D^2 \times \{q\}$, for some $q \in S^1$,
rotating one of the disks through an angle $\frac{m}{n}\,2\,\pi$,
where $m$, $n$ are integers, and gluing them
back again. While the central fiber now still closes after one cycle,
the remaining fibers close in general
only after $n$ cycles. We note that there exists a fiber preserving $n
: 1$ map $\widehat{\psi}$ of the ordinary fibered solid torus onto the
fibered solid torus which is a local diffeomorphism that induces a $n
: 1$ covering map on the central fiber.

As shown in \cite{im2}, it follows from Epstein's theorem \cite{ep} that the 
null geodesics on $\mathcal{N}$ represent the fibers of a Seifert fibration. 
This means that any closed null geodesic has a ``fibered neighborhood'', i.e.
a neighborhood fibered by closed null geodesics, which can be mapped by a fiber
preserving diffeomorphism onto a fibered solid torus.
Because $\mathcal{N}$ is compact, it can be covered by a finite number of such
fibered neighborhoods, $\mathcal{N}_i$. 
For any neighborhood $\mathcal{N}_i$ there is an ordinary fibered solid 
torus $\widehat{\mathcal{N}}_i$ which is mapped by a fiber preserving map
$\widehat{\psi}_i$ onto $\mathcal{N}_i$ as described above. Further we can choose
a tubular spacetime neighborhood, ${\cal U}_i$, of $\mathcal{N}_i$, so that 
${\cal U}_i$ has  topology $D^2  \times \mathbb{R} \times S^1$ and the fibration
of $\mathcal{N}_i$ extends to ${\cal U}_i$. There exists then a fibered extension
$\widehat{{\cal U}}_i \simeq D^2  \times \mathbb{R}\times S^1$ of 
$\widehat{\mathcal{N}}_i$,
with fibers $\{p\} \times S^1$, $p \in D^2  \times \mathbb{R}$, to which 
$\widehat{\psi}_i$ can be extended to a fiber preserving local diffeomorphism
which maps  $\widehat{\cal U}_i$ onto ${\cal U}_i$. We denote the extension again by
$\widehat{\psi}_i$. Let
$\mathcal{O}_i$ denote the universal covering space of $\widehat{\cal U}_i$ (so that
$\mathcal{O}_i$ has topology $D^2 \times \mathbb{R}^2$). We denote the
projection map from
$\mathcal{O}_i$ onto $\widehat{\cal U}_i$ by $\widetilde{\psi}_i$, set
$\psi_i = \widehat{\psi}_i \circ \widetilde{\psi}_i$, and denote the inverse
image of $\mathcal{N}_i$ under $\psi_i$ by $\widetilde{\mathcal{N}}_i$.

We will refer to $(\mathcal{O}_i, \psi^{*}_ig_{ab})$ as an {\em elementary
spacetime region}. Note that for the case of a spacetime of class B constructed
from a spacetime of class A in the manner of Proposition \ref{AB},
$\mathcal{O}_i$ may be identified with a neighborhood of a portion of
the horizon in the original (class A) spacetime.

Our main results will be based upon the following theorem, which may
be extracted directly from the analysis of Isenberg and Moncrief
\cite{im1,im2}:

\begin{theorem}\label{theor-mi} 
({Moncrief $\&$ Isenberg}) Let $(M,g_{ab})$ be a smooth electrovac
spacetime of class B and let $(\mathcal{O}_i, \psi^{*}_ig_{ab})$ be an
elementary spacetime region, as defined above. Then, there exists a Gaussian 
null coordinate system $(u,r,x^3,x^4)$ (see Appendix A) covering a
neighborhood, $\mathcal{O}_i'$, of $\widetilde{\mathcal{N}}_i$ 
in $\mathcal{O}_i$ so that the following properties hold 
(i) The coordinate range of $u$ is $-\infty < u < \infty$ whereas the coordinate
range of $r$ is $-\epsilon < r < \epsilon$ for some $\epsilon > 0$, with
the surface $r = 0$ being $\widetilde{\mathcal{N}}_i$. 
(ii) In $\mathcal{O}_i'$, the projection map 
$\widetilde{\psi}_i : \mathcal{O}_i \rightarrow \widehat{\cal U}_i$ 
is obtained by periodically identifying the
coordinate $u$ with some period $P \in \mathbb{R}$. Thus, in particular, 
the components of $\psi^{*}_ig_{ab}$ and $\psi^{*}_iF_{ab}$ in these
coordinates are periodic functions of $u$ with period $P$. 
(iii) We have, writing in the following for convenience
$g_{ab}$ and $F_{ab}$ instead of $\psi^{*}_ig_{ab}$ and
$\psi^{*}_iF_{ab}$,
\begin{equation}
\left.f\right| _{\widetilde{\mathcal{N}}_i}=-2\kappa_{\circ}\ \ {\rm and}\ \ 
\left.F_{uA}\right| _{\widetilde{\mathcal{N}}_i}=0,   \label{f0}
\end{equation}
with $\kappa_{\circ}\in\mathbb{R}$, where $f$ is defined in Appendix A.
(iv) On
$\widetilde{\mathcal{N}}_i$, the $r$-derivatives of the metric and
Maxwell field tensor components up to any order are $u$-independent,
i.e., in the notation of Appendix A
\begin{equation}
\left.\frac{\partial}{\partial u} \left[\frac{\partial ^n}{\partial r^n}
\left\{f,h_A,g_{AB};F_{ur},F_{uA},F_{rA},F_{AB}\right\}
\right] \right| _{\widetilde{\mathcal{N}}_i}=0,\label{diff}
\end{equation}
for all $n\in\mathbb{N}\cup\{0\}$.
\end{theorem}

\begin{remark} \label{kappa}
Along the null geodesic generators of $\widetilde{\mathcal{N}}_i$
the vector field $k^{a}=\left(\partial/\partial u\right)^{a}$
(which we take to be future directed) satisfies the equation
\begin{equation}
k^{a}\nabla _{a}k^{b}=\kappa_{\circ} k^{b}.\label{geod}
\end{equation}
Thus, if $\kappa _{\circ } > 0$, then all of the null geodesic
generators of $\widetilde{\mathcal{N}}_i$ are past incomplete but
future complete. If $\kappa _{\circ } < 0$ then all of the null
geodesic generators of $\widetilde{\mathcal{N}}_i$ are past complete
but future incomplete. Similarly, if $\kappa _{\circ } = 0$ (usually
referred to as the ``degenerate case''), then $u$ is an affine
parameter and all of the null geodesic generators of
$\widetilde{\mathcal{N}}_i$ are complete in both the past and future
directions.
\end{remark}

\begin{remark} \label{proj}
In the analytic case, eq.(\ref{diff}) directly implies that
$k^a=(\partial/\partial u)^a$ is a Killing vector field in a
neighborhood of $\widetilde{\mathcal{N}}_i$. Since the projection map
$\widetilde{\psi}_i$ is obtained by periodically identifying the
coordinate $u$, it follows immediately that $k^a$ projects to a
Killing vector field $\widehat{k}^a$ in a neighborhood of
$\widehat{\mathcal{N}}_i$.  Appealing then to the argument of
\cite{im2}, it can be shown that the $\widehat{k}^a$ further projects
to a Killing vector field under the action of the map
$\widehat{\psi}_i$, so that the map
$\psi_i=\widetilde{\psi}_i\circ\widehat{\psi}_i$ projects
$k^a=(\partial/\partial u)^a$ to a well-defined Killing field in a
neighborhood of ${\mathcal{N}}_i$.  The arguments of \cite{im1,im2}
then establish that the local Killing fields
obtained for each fibered neighborhood can be patched together to
produce a global Killing field on a neighborhood of $\mathcal{N}$.
\end{remark}

In the next section, we shall generalize the result of Remark
\ref{proj} to the smooth case. However, to do so we will need to
impose the additional restriction that $\kappa_{\circ} \neq 0$, and we
will prove existence only on a one-sided neighborhood of the horizon.

\section{Existence of a Killing vector field}
\setcounter{equation}{0}

The main difficulty encounted when one attempts to generalize the
Isenberg-Moncrief theorem to the smooth case is that suitable detailed
information about the spacetime metric and Maxwell field is known only
on $\mathcal{N}$ (see eq.(\ref{diff}) above).  If a Killing field
$k^a$ exists, it is determined uniquely by the data $k_a$,
$\nabla_{[a}\,k_{b]}$ at one point of $\mathcal{N}$, because equations
(\ref{Killing}), (\ref{Kint}) imply a system of ODE's for the tetrad
components $k_j$, $\nabla_{[i}\,k_{j]}$ along each $C^1$ curve. But
the existence of a Killing field cannot be shown this way. Thus we
will construct the Killing field as a solution to a PDE problem.
However, $\mathcal{N}$ is a null surface, and thus, by itself, it does
not comprise a suitable initial data surface for the relevant
hyperbolic equations. We now remedy this difficulty by performing a
suitable local extension of a neighborhood of
$\widetilde{\mathcal{N}}_i$ which is covered by the Gaussian null
coordinates of Theorem \ref{theor-mi}. This is achieved via the
following Proposition:

\begin{proposition}\label{theor-ext}
Let $(\mathcal{O}_i,g_{ab}\mid _{\mathcal{O}_i})$ be an elementary
spacetime region associated with an electrovac spacetime of class B
such that $\kappa_{\circ}>0$ (see eq.(\ref{f0}) above).  Then, there
exists an open neighborhood, $\mathcal{O}_i''$, of
$\widetilde{\mathcal{N}}_i$ in $\mathcal{O}_i$ such that
$(\mathcal{O}_i'',g_{ab}\mid _{\mathcal{O}_i''},F_{ab}\mid
_{\mathcal{O}_i''})$ can be extended to a smooth electrovac spacetime,
$(\mathcal{O}^*,g^*_{ab},F^*_{ab})$, that possesses a bifurcate null
surface, $\widetilde{\mathcal{N}}^*$---i.e.,
$\widetilde{\mathcal{N}}^*$ is the union of two null hypersurfaces,
$\mathcal{N}^*_1$ and $\mathcal{N}^*_2$, which intersect on a
2-dimensional spacelike surface, $S$---such that
$\widetilde{\mathcal{N}}_i$ corresponds to the portion of
$\mathcal{N}^*_1$ that lies to the future of $S$ and $I^+[S] =
\mathcal{O}_i'' \cap I^+[\widetilde{\mathcal{N}}_i]$. Furthermore, the
expansion and shear of both $\mathcal{N}^*_1$ and $\mathcal{N}^*_2$
vanish.
\end{proposition}

\noindent{\bf Proof}{\ } It follows from eq.(\ref{diff}) that in
$\mathcal{O}_i'$, the spacetime metric $g_{ab}$ can be decomposed as
\begin{equation}
g_{ab} = g^{(0)}_{ab} + \gamma_{ab}
\end{equation}
where, in the Gaussian null coordinates of Theorem \ref{theor-mi}, the
components, $g^{(0)}_{\mu \nu}$, of $g^{(0)}_{ab}$ are independent of
$u$, whereas the components, $\gamma_{\mu \nu}$, of $\gamma_{ab}$ and
all of their derivatives with respect to $r$ vanish at $r=0$ (i.e., on
$\widetilde{\mathcal{N}}_i$). Furthermore, taking account of the
periodicity of $\gamma_{\mu \nu}$ in $u$ (so that, in effect, the
coordinates $(u,x^3,x^4)$ have a compact range of variation), we see
that for all integers $j \geq 0$ we have throughout $\mathcal{O}_i'$
\begin{equation}
|\gamma_{\mu \nu}| < C_j |r|^j
\label{gammabnd}
\end{equation}
for some constants $C_j$. Similar relations hold for all partial
derivatives of $\gamma_{\mu \nu}$.

It follows from eq.(\ref{gammabnd}) that there is an open neighborhood
of $\widetilde{\mathcal{N}}_i$ in the spacetime
$(\mathcal{O}_i',g^{(0)}_{ab})$ such that $g^{(0)}_{ab}$ defines a
Lorentz metric. It is obvious that in this neighborhood,
$\widetilde{\mathcal{N}}_i$ is a Killing horizon of $g^{(0)}_{ab}$
with respect to the Killing field $k^a = (\partial / \partial
u)^a$. Consequently, by the results of \cite{rw1,rw2}, we may extend
an open neighborhood, $\mathcal{O}_i''$, of
$\widetilde{\mathcal{N}}_i$ in $\mathcal{O}_i$ to a smooth spacetime
$(\mathcal{O}^*,g^{(0)*}_{ab})$, that possesses a bifurcate Killing
horizon, $\widetilde{\mathcal{N}}^*$, with respect to
$g^{(0)*}_{ab}$. Furthermore, with respect to the metric
$g^{(0)*}_{ab}$, $\widetilde{\mathcal{N}}^*$ automatically satisfies
all of the properties stated in the Proposition. In addition, by
theorem 4.2 of \cite{rw2}, the extension can be chosen so that $k^a$
extends to a Killing field, $k^{*a}$, of $g^{(0)*}_{ab}$ in
$\mathcal{O}^*$, and $(\mathcal{O}^*,g^{(0)*}_{ab})$ possesses a
``wedge reflection'' isometry (see \cite{rw2}); we assume that such a
choice of extension has been made.

Let $(u_0,r_0,x^3_0,x^4_0)$ denote the Gaussian null coordinates in
$\mathcal{O}_i''$ associated with $g^{(0)}_{ab}$, such that on
$\widetilde{\mathcal{N}}_i$ we have $r_0 = r = 0, u_0 = u, x^3_0 =
x^3, x^4_0 = x^4$. Since $\gamma_{ab}$ is smooth in $\mathcal{O}_i''$
and is periodic in $u$, it follows that each of the coordinates
$(u_0,r_0,x^3_0,x^4_0)$ are smooth functions of $(u,r,x^3,x^4)$ which
are periodic in $u$. It further follows that the Jacobian matrix of the
transformation between $(u_0,r_0,x^3_0,x^4_0)$ and $(u,r,x^3,x^4)$ is
uniformly bounded in $\mathcal{O}_i''$, and that, in addition, there exists a
constant, $c$ such that $|r| \leq c |r_0|$ in
$\mathcal{O}_i''$. Consequently, the components, $\gamma_{\mu_0
\nu_0}$ of $\gamma_{ab}$ in the Gaussian null coordinates associated
with $g^{(0)}_{ab}$ satisfy for all integers $j \geq 0$
\begin{equation}
|\gamma_{\mu_0 \nu_0}| < C_j' |r_0|^j
\label{gamma0bnd}
\end{equation}

Let $(U,V)$ denote the generalized Kruskal coordinates with respect to
$g^{(0)}_{ab}$ introduced in \cite{rw1,rw2}. In terms of these
coordinates, $\mathcal{O}_i''$ corresponds to the portion of
$\mathcal{O}^*$ satisfying $U > 0$ and the wedge reflection isometry
mentioned above is given by $U \rightarrow -U, V \rightarrow - V$.
The null hypersurfaces $\mathcal{N}^*_1$ and $\mathcal{N}^*_2$ that
comprise the bifurcate Killing horizon, $\widetilde{\mathcal{N}}^*$,
of $g^{(0)*}_{ab}$ correspond to the hypersurfaces defined by $V = 0$
and $U = 0$, respectively.

It follows from eq.(23) of \cite{rw1} that within $\mathcal{O}_i''$
we have
\begin{equation}
|r_0| < C |UV|
\end{equation}
for some constant $C$. Hence, we obtain for all $j$
\begin{equation}
|\gamma_{\mu_0 \nu_0}| < C_j'' |UV|^j
\end{equation}
with similar relations holding for all of the derivatives of
$\gamma_{\mu_0 \nu_0}$ with respect to the coordinates
$(u_0,r_0,x^3_0,x^4_0)$. Taking account of the transformation between
the Gaussian null coordinates and the generalized Kruskal coordinates
(see eqs. (24) and (25) of \cite{rw1}), we see that the Kruskal
components of $\gamma_{ab}$ and all of their Kruskal coordinate
derivatives also go to zero uniformly on compact subsets of
$(V,x^3,x^4)$ in the limit as $U \rightarrow 0$. It follows that the
tensor field $\gamma_{ab}$ on $\mathcal{O}_i''$ extends smoothly to $U
= 0$---i.e., the null hypersurface $\mathcal{N}^*_2$ of
$\mathcal{O}^*$---such that $\gamma_{ab}$ and all of its derivatives
vanish on $\mathcal{N}^*_2$. We now further extend $\gamma_{ab}$ to
the region $U < 0$---thereby defining a smooth tensor field
$\gamma^*_{ab}$ on all of $\mathcal{O}^*$---by requiring it to be
invariant under the above wedge reflection isometry. In
$\mathcal{O}^*$, we define
\begin{equation}
g^*_{ab} = g^{(0)*}_{ab} + \gamma^*_{ab}
\end{equation}
Then $g^*_{ab}$ is smooth in $\mathcal{O}^*$ and is invariant under
the wedge reflection isometry. Furthermore, since $\gamma^*_{ab}$
vanishes on $\widetilde{\mathcal{N}}^*$, it follows that
$\widetilde{\mathcal{N}}^*$ is a bifurcate null surface with respect
to $g^*_{ab}$ and that $k^{*a}$ is normal to
$\widetilde{\mathcal{N}}^*$. In addition, on
$\widetilde{\mathcal{N}}^*$ we have ${\cal L}_{k^*} g^{(0)*}_{ab} = 0$
(since $k^{*a}$ is a Killing field of $g^{(0)*}_{ab}$) and ${\cal
L}_{k^*} \gamma^*_{ab} = 0$ (since $\gamma^*_{ab}$ and its derivatives
vanish on $\widetilde{\mathcal{N}}^*$). Therefore, we have ${\cal
L}_{k^*} g^*_{ab} = 0$ on
$\widetilde{\mathcal{N}}^*$. By Lemma \ref{lem-rs=0}, it follows that the
expansion and shear of both $\mathcal{N}^*_1$ and $\mathcal{N}^*_2$
vanish.

Finally, by a similar construction (using the fact that 
$F_{uA}\vert_{\widetilde{\mathcal{N}}_i}=0$; see eq.(\ref{f0})), 
we can extend the Maxwell field $F_{ab}$ in
$\mathcal{O}_i''$ to a smooth Maxwell field $F^*_{ab}$ in
$\mathcal{O}^*$ which is invariant under the wedge reflection
isometry. By hypothesis, $(g^*_{ab}, F^*_{ab})$ satisfies the
Einstein-Maxwell equations in the region $U > 0$. By invariance under
the wedge reflection isometry, $(g^*_{ab}, F^*_{ab})$ also
satisfies the Einstein-Maxwell equations in the region $U < 0$. By
continuity, the Einstein-Maxwell equations also are satisfied for $U =
0$, so $(g^*_{ab}, F^*_{ab})$ is a solution throughout
$\mathcal{O}^*$.   \hfill\fbox{}

\begin{remark}
By remark \ref{kappa}, the hypothesis that $\kappa _{\circ } > 0$ is
equivalent to the condition that the null geodesic generators of
$\widetilde{\mathcal{N}}_i$ are past incomplete. Therefore, it is
clear that Proposition \ref{theor-ext} also holds for $\kappa _{\circ
} < 0$ if we interchange futures and pasts. However, no analog of
Proposition \ref{theor-ext} holds for the ``degenerate case'' $\kappa
_{\circ } = 0$.
\end{remark}

We are now prepared to state and prove our main theorem:

\begin{theorem}
Let $(M,g_{ab})$ be a smooth electrovac spacetime of class B for which
the generators of the null hypersurface $\mathcal{N}$ are past
incomplete. Then there exists an open neighborhood, $\mathcal{V}$ of
$\mathcal{N}$ such that in $J^+[\mathcal{N}] \cap \mathcal{V}$ there
exists a smooth Killing vector field $k^a$ which is normal to
$\mathcal{N}$. Furthermore, in $J^+[\mathcal{N}] \cap \mathcal{V}$ the
electromagnetic field, $F_{ab}$, satisfies $\mathcal{L}_kF_{ab}=0$.
\end{theorem}

\noindent{\bf Proof}{\ } As explained in Section 3, we can cover
$\mathcal{N}$ by a finite number of fibered neighborhoods,
$\mathcal{N}_i$. Let $\mathcal{O}_i$ denote the elementary spacetime
region obtained by ``unwrapping'' a neighborhood of $\mathcal{N}_i$,
as explained in Section 3. By remark \ref{kappa}, the past
incompleteness of the null geodesic generators of $\mathcal{N}$
implies that $\kappa _{\circ } > 0$, so Proposition \ref{theor-ext}
holds. We now apply Proposition \ref{Kex} to the extended spacetime
$\mathcal{O}^*$ to obtain existence of a Killing vector field in the
domain of dependence of $\widetilde{\mathcal{N}}^* = \mathcal{N}^*_1
\cup \mathcal{N}^*_2$. By restriction to $\mathcal{O}_i$, we thereby
obtain a Killing field $K^a$ (which also Lie derives the Maxwell
field) on a one-sided neighborhood of $\widetilde{\mathcal{N}}_i$ of
the form $J^+[\widetilde{\mathcal{N}}_i] \cap
\widetilde{\mathcal{V}}_i$, where $\widetilde{\mathcal{V}}_i$ is an
open neighborhood of $\widetilde{\mathcal{N}}_i$. Both $K^a$ and
$k^a=(\partial/\partial u)^a$ are tangent to the null geodesic
generators of $\widetilde{\mathcal{N}}_i$, so on
$\widetilde{\mathcal{N}}_i$ we clearly have $K^a = \varphi k^a$ for
some function $\varphi$. Furthermore, on $\widetilde{\mathcal{N}}_i$,
we have $\mathcal{L}_Kg_{ab} = 0$ (since $K^a$ is a Killing field) and
$\mathcal{L}_kg_{ab} = 0$ (as noted in the proof of Proposition
4.1). It follows immediately that $\nabla_a \varphi = 0$ on
$\widetilde{\mathcal{N}}_i$, so we may rescale $K^a$ so that $K^a =
k^a$ on $\widetilde{\mathcal{N}}_i$. Since the construction of $k^a$
off of $\widetilde{\mathcal{N}}_i$ (as described in Appendix A) is
identical to that which must be satisfied by a Killing field (as
described in Remark B.1 below), it follows that $K^a = k^a$ within
their common domain of definition. Thus, the vector field
$k^a=(\partial/\partial u)^a$---which previously had been shown to be
a Killing field in the analytic case---also is a Killing field in the
smooth case in $J^+[\widetilde{\mathcal{N}}_i] \cap
\widetilde{\mathcal{V}}_i$.  By exactly the same arguments as given in
\cite{im1,im2} (see Remark \ref{proj} above) it then follows that the
map $\psi_i=\widetilde{\psi}_i\circ\widehat{\psi}_i$ projects
$k^a=(\partial/\partial u)^a$ to a well-defined Killing field in a
one-sided neighborhood of ${\mathcal{N}}_i$, and that the local
Killing fields obtained for each fibered neighborhood can be patched
together to produce a global Killing field on a one-sided neighborhood
of $\mathcal{N}$ of the form $J^+[\mathcal{N}]\cap\mathcal{V}$ where
$\mathcal{V}=\cup_i\psi_i[\mathcal{V}_i]$.  \hfill\fbox{}

\medskip

In view of Proposition \ref{AB}, we have the following Corollary

\begin{corollary}
Let $(M,g_{ab})$ be a smooth electrovac spacetime of class A for which
the generators of the event horizon $\mathcal{N}$ are past
incomplete. Then there exists an open neighborhood, $\mathcal{V}$ of
$\mathcal{N}$ such that in $J^+[\mathcal{N}] \cap \mathcal{V}$ there
exists a smooth Killing vector field $k^a$ which is normal to
$\mathcal{N}$. Furthermore, in $J^+[\mathcal{N}] \cap \mathcal{V}$ the
electromagnetic field, $F_{ab}$, satisfies $\mathcal{L}_kF_{ab}=0$.
\end{corollary}

\section*{Acknowledgments}

This research was supported in parts by Monbusho Grant-in-aid
No. 96369 and by NSF grant PHY 95-14726 to the University of
Chicago. We wish to thank Piotr Chrusciel and James Isenberg for
reading the manuscript. One of us (IR) wishes to thank the Albert
Einstein Institute and the Physics Department of the Tokyo Institute
of Technology for their kind hospitality during part of the work on
the subject of the present paper.

\appendix

\section{Gaussian null coordinate systems}
\setcounter{equation}{0}

In this Appendix the construction of a local Gaussian
null coordinate systems will be recalled. 

\medskip

Let $(M,g_{ab})$ be a spacetime, let $\mathcal{N}$ be a smooth null
hypersurface, and let $\varsigma$ be a smooth spacelike 2-surface
lying in $\mathcal{N}$.  Let $\left(x^3,x^4\right)$ be coordinates on
an open subset $\widetilde\varsigma$ of $\varsigma $. On a
neighborhood of $\widetilde\varsigma$ in $\mathcal{N}$, let $k^a$
be a smooth, non-vanishing normal vector field to $\mathcal{N}$, so
that the integral curves of $k^a$ are the null geodesic generators of
$\mathcal{N}$. Without loss of generality, we may assume that $k^a$ is
future directed.

On a sufficiently small open neighborhood, $S$, of
$\widetilde\varsigma \times \left\{ 0\right\}$ in $\widetilde\varsigma
\times \mathbb{R}$, let $\psi :S \rightarrow \mathcal{N}$ be the map
which takes $\left( q,u\right)$ into the point of $\mathcal{N}$ lying
at parameter value $u$ along the integral curve of $k^a$ starting at
$q$.  Then, $\psi$ is $C^\infty$, and it follows from the inverse
function theorem that $\psi $ is $1:1$ and onto from an open
neighborhood of $\widetilde\varsigma \times \left\{ 0\right\}$ onto an
open neighborhood, $\widetilde{\mathcal{N}}$, of $\widetilde\varsigma
$ in $ \mathcal{N}$. Extend the functions $x^3,x^4$ from
$\widetilde\varsigma $ onto $\widetilde{\mathcal{N}}$ by keeping their
values constant along the integral curves of $k^a$. Then $\left(
u,x^3,x^4\right) $ are coordinates on $\widetilde{\mathcal{N}}$.

At each $p\in \widetilde{\mathcal{N}}$ let $l^a$ be the unique null
vector field on $\widetilde{\mathcal{N}}$ satisfying $l^ak_a=1$ and
$l^aX_a=0$ for all $X^a$ which are tangent to $\widetilde{
\mathcal{N}}$ and satisfy $X^a\nabla _au=0$. On a sufficiently small
open neighborhood, $Q$, of $\widetilde{\mathcal{N}}\times \left\{
0\right\} $ in $\widetilde{\mathcal{N}}\times \mathbb{R}$, let $\Psi
:Q \rightarrow M$ be the map which takes $\left( p,r\right) \in Q$
into the point of $M$ lying at affine parameter value $r$ along the
null geodesic starting at $p$ with tangent $l^a$. Then $\Psi $ is
$C^\infty $ and it follows from the inverse function theorem
that $\Psi $ is $1:1$ and onto from an open
neighborhood of $\widetilde{\mathcal{N}}\times \left\{ 0\right\} $
onto an open neighborhood, $\mathcal{O}$, of
$\widetilde{\mathcal{N}}$ in $M.$ We extend the functions $u,x^3,x^4$
from $\widetilde{\mathcal{N}}$ to $\mathcal{O}$ by requiring their
values to be constant along each null geodesic determined by
$l^a$. Then $\left(u,r,x^3,x^4\right) $ yields a coordinate system on
$\mathcal{O}$ which will be referred to as \textit{Gaussian null}
coordinate system. Note that on $\widetilde{\mathcal{N}}$ we have
$k^a=\left(\partial /\partial u\right)^a$.

Since by construction the vector field $l^a=\left( \partial /\partial
r\right) ^a$ is everywhere tangent to null geodesics we have that
$g_{rr}=0$ throughout $\mathcal{O}$. Furthermore, we have that the
metric functions $g_{ru},g_{r3},g_{r4}$ are independent of $r$, i.e.
$g_{ru}=1,g_{r3}=g_{r4}=0$ throughout $\mathcal{O}$.  In addition, as
a direct consequence of the above construction, $g_{uu}$ and $g_{uA}$
vanish on $\widetilde{\mathcal{N}}$.  Hence, within $\mathcal{O}$,
there exist smooth functions $f$ and $h_A$, with
$f\vert_{\widetilde{\mathcal{N}}}=(\partial g_{uu}/\partial r)\mid
_{r=0}$ and $h_A\vert_{\widetilde{\mathcal{N}}}=(\partial
g_{uA}/\partial r)\mid_{r=0}$, so that the spacetime metric in
$\mathcal{O}$ takes the form
\begin{equation}
\mathrm{d}s^2=r\cdot f\mathrm{d}u^2+2\mathrm{d}r\mathrm{d}u+2r\cdot
h_{A}\mathrm{d}u \mathrm{d}x^A+g_{AB}\mathrm{d}x^A\mathrm{d}x^B
\label{le1}
\end{equation}
where $g_{AB}$ are smooth functions of $u,r,x^3,x^4$ in $\mathcal{O}$
such that $g_{AB}$ is a negative definite $2\times 2$ matrix, and the
uppercase Latin indices take the values $3,4$.

\section{The existence of a Killing field tangent to the horizon}
\setcounter{equation}{0}

The purpose of this section is to prove the following fact.

\begin{proposition} 
\label{Kex}
Suppose that $(M, g, F)$ is a time oriented solution to the
Einstein-Maxwell equations with Maxwell field $F$ (without
sources). Let ${\cal N}_1$, ${\cal N}_2$ be smooth null hypersurfaces
with connected space-like boundary ${\cal Z}$, smoothly embedded in
$M$, which are generated by the future directed null geodesics
orthogonal to ${\cal Z}$. Assume that ${\cal N}_1 \cup {\cal N}_2$ is
achronal. Then there exists on the future domain of dependence, $D^{+}$, of 
${\cal N}_1 \cup {\cal N}_2$ a non-trivial Killing field $K$ which is tangent 
to the null generators of ${\cal N}_1$ and ${\cal N}_2$ if and only if these 
null hypersurfaces are expansion and shear free. If it exists, the Killing
field is unique up to a constant factor and we have ${\cal L}_K\,F = 0$.
\end{proposition}

\begin{remark}
We shall prove Proposition \ref{Kex} by first deducing the form that
$K$ must have on ${\cal N}_1 \cup {\cal N}_2$, then defining $K$ on
$D^+$ by evolution from ${\cal N}_1 \cup {\cal N}_2$ of a wave
equation that must be satisfied by a Killing field, and, finally,
proving that the resulting $K$ is indeed a Killing field. An
alternative, more geometric, approach to the proof of Proposition
\ref{Kex} would be to proceed as follows. Suppose that the Killing
field $K$ of Proposition \ref{Kex} exists. Denote by $\psi_t$ the
local 1-parameter group of isometries associated with $K$. Since
$\psi_t$ maps ${\cal N}_1 \cup {\cal N}_2$ into itself, geodesics into
geodesics, and preserves affine parameterization, we can describe for
given $t$ the action of $\psi_t$ on a neighborhood of ${\cal N}_1 \cup
{\cal N}_2$ in $D^{+}$ in terms of its action on geodesics passing
through ${\cal N}_1 \cup {\cal N}_2$ into $D^{+}$ and affine
parameters which vanish on ${\cal N}_1 \cup {\cal N}_2$. There are
various possibilities, one could employ e.g. the future directed
time-like geodesics starting on ${\cal Z}$ or the null geodesics which
generate double null coordinates adapted to ${\cal N}_1 \cup {\cal
N}_2$. Changing the point of view, one could try to use such a
description to define maps $\psi_t$, show that they define a local
group of isometries, and then define $K$ as the corresponding Killing
field. To show that $\psi_t^{*}\,g = g$, one would prove this relation
on ${\cal N}_1 \cup {\cal N}_2$ and then invoke the uniqueness for the
characteristic initial value problem for the Einstein-Maxwell
equations to show that this relation holds also in a neighborhood of
${\cal N}_1 \cup {\cal N}_2$ in $D^{+}$. For this to work we would
need to show that $\psi_t^{*}\,g$ has a certain smoothness ($C^2$
say). This is not so difficult away from ${\cal N}_1 \cup {\cal N}_2$
but it is delicate near the initial hypersurface. The discussion would
need to take into account the properties of the underlying space-time
exhibited in Lemmas \ref{lem-rs=0}, \ref{lem-intdata} below and would
become quite tedious. For this reason we have chosen not to proceed in
this manner.

\end{remark}

It will be convenient to use the formalism, notation, and conventions
of \cite{newman:penrose} in a gauge adapted to our geometrical
situation. Since we will be using the tetrad formalism, throughout
this Appendix we shall omit all abstract indices $a,b,c...$ on
tensors and use the indices $i,j,k...$ to denote the components
of tensors in our tetrad. We begin by choosing smooth coordinates
$u = x^1$, $r = x^2$, $x^A$, $A = 3, 4$, and a smooth tetrad field
\[
z_1 = l,\,\,z_2 = n\,\,z_3 = m,\,\,z_4 = \overline{m},
\]
with $g_{ik} = g(z_i, z_k)$ such that $g_{12} = g_{21} = 1$,
$g_{34} = g_{43} = -1$ are the only non-vanishing scalar products.
Let $x^A$ be coordinates on a connected open
subset $\zeta$ of ${\cal Z}$ on which also $m$, $\overline{m}$ can be
introduced such that they are tangent to $\zeta$. On $\zeta$ we set $x^1 = 0$,
$x^2 = 0$ and assume that $l$ is tangent to ${\cal N}_1$, $n$ is tangent to
${\cal N}_2$, and both are future directed. The possible choices which can be
made above will represent the remaining freedom in our gauge. We assume   
\[
\nabla_{z_2}\,z_2 = 0,\,\,\,\,\,<z_2, d\,x^{\mu}>\, = \delta^{\mu}\,_1, 
\quad\mbox{on}\quad N'_2,      
\]
and set $\zeta_c = \{x^1 = c\} \subset {\cal N}'_2$ for $c \ge 0$, where
${\cal N}'_2$ denotes the subset of ${\cal N}_2$ generated by the null
geodesics starting on $\zeta$. We assume that $m$, $\overline{m}$ are tangent
to $\zeta_c$. From the transformation law of the spin coefficient
$\gamma$ under rotations $m \rightarrow e^{i\,\phi}\,m$ we find that we can
always assume that $\gamma = \overline{\gamma}$ on ${\cal N}'_2$. With this
assumption $m$, whence also $l$, will be fixed uniquely on ${\cal N}'_2$ and
we have  
\[
\gamma = 0,\,\,\,\,\,\nu = 0 \quad\mbox{on}\quad {\cal N}'_2.
\] 
The coordinates and the frame are extended off ${\cal N}'_2$ such that
\[
\nabla_{z_1}\,z_i = 0,\,\,\,\,\,<z_1, d\,x^{\mu}>\, = \delta^{\mu}\,_2. 
\]
On a certain neighborhood, $D$, of ${\cal N}'_1 \cup {\cal N}'_2$ in $D^{+}$
(where ${\cal N}'_1$ is the subset of ${\cal N}_1$ generated by the null
geodesics starting on $\zeta$), we obtain by this procedure a smooth coordinate
system and a smooth frame field which has in these coordinates the local
expression  
\[
l_{\mu} = \delta^1\,_{\mu},\,\,\,\,\, 
l^{\mu} = \delta^{\mu}\,_2,\,\,\,\,\, 
n^{\mu} = \delta^{\mu}\,_1 + U\,\delta^{\mu}\,_2 + X^A\,\delta^{\mu}\,_A,
\,\,\,\,\,
m^{\mu} = \omega\,\delta^{\mu}\,_2 + \xi^A\,\delta^{\mu}\,_A.
\]
We have ${\cal N}'_1 = \{x^1 = u = 0\}$, ${\cal N}'_2 = \{x^2 = r = 0\}$
and
\[
\kappa = 0,\,\,\,\,\epsilon = 0,\,\,\,\,\pi = 0,\,\,\,\,
\tau = \overline{\alpha} + \beta \quad\mbox{on}\quad D,\,\,\,\,\,
U = 0,\,\,\,\,X^A = 0,\,\,\,\,\omega = 0 \quad\mbox{on}\quad {\cal N}'_2.
\]
We shall use alternatively the Ricci rotation coefficients defined by 
$\nabla_i\,z_j \equiv \nabla_{z_i}\,z_j = \gamma_j\,^l\,_i\,z_l$ or their
representation in terms of spin coefficients as given in
\cite{newman:penrose}. The gauge above will be used in many local
considerations whose results  extend immediately to all of ${\cal N}_1$ and
${\cal N}_2$. We shall then always state the extended result.  

We begin by showing the necessity of the conditions on the null hypersurfaces
in Proposition \ref{Kex} and some of their consequences.

\begin{lemma}
\label{lem-rs=0}
Let ${\cal N}$ be a smooth null hypersurface of the space-time $(M, g)$ and
$X$ a smooth vector field on $M$ which is tangent to the null generators of
${\cal N}$ and does not vanish there. If $g'$ denotes the pull back of $g$ to
${\cal N}$, then  ${\cal L}_X\,g' = 0$ on ${\cal N}$ if and only if the null
generators of ${\cal N}$ are expansion and shear free.   
\end{lemma}

\noindent{\bf Proof}{\ } We can assume that ${\cal N}$ coincides with
the hypersurface ${\cal N}_1$.  Then we have in our gauge $X =
X^1\,z_1$ on ${\cal N}$ with $X^1 \neq 0$, and ${\cal L}_X\,g' = 0$
translates into $0 = \nabla_{(i}\,X_{j)} =
z_{(i}(X^1)\,\delta^2\,_{j)} - \gamma_{(j}\,^2\,_{i)}\,X^1$ on ${\cal
N}$ with $i, j \neq 2$. By our gauge this is equivalent to $0 =
\gamma_{(A}\,^2\,_{B)} = - \sigma\,\delta^3\,_A\,\delta^3\,_B -
Re\,\rho\,\,\delta^3\,_{(A}\,\delta^4\,_{B)} -
\bar{\sigma}\,\delta^4\,_A\,\delta^4\,_B$. Since $\rho$ is real on the
hypersurface ${\cal N}$, the assertion follows. \hfill\fbox{}

\begin{lemma}
\label{lem-intdata}
If the null hypersurfaces ${\cal N}_1$, ${\cal N}_2$ are expansion and
shear free, then the frame coefficients, the spin coefficients, the
components $\Psi_i$ of the conformal Weyl spinor field, the components
$\phi_k$ of the Maxwell spinor field, and the components $\Phi_{ik} =
k\,\phi_i\,\bar{\phi}_k$ of the Ricci spinor field are uniquely
determined in our gauge on ${\cal N}'_1$ and ${\cal N}'_2$ by
the field equations and the data
\begin{equation}
\label{Zdata}
\phi_1,\,\,\,\,\,\tau,\,\,\,\,\,\xi^A, \,\,\,A = 3, 4,
\quad\mbox{on}\quad {\cal Z}.
\end{equation}
In particular, we have 
\[
\Psi_0 = 0,\,\,\Psi_1 = 0,\,\,\phi_0 = 0,\,\,
\Phi_{0k} = \overline{\Phi}_{k0} = 0,
\]
\begin{equation}
\label{N1dat}
D\,\phi_1 = 0,\,\,
D\,\phi_2 = \overline{\delta}\,\phi_1,\,\,
\phi_2 = r\,\overline{\delta}\,\phi_1,
\,\,\omega = - r\,\tau,\,\,\mu = r\,\Psi_2,
\quad\mbox{on}\quad {\cal N}'_1,
\end{equation}  

\[
\Psi_4 = 0,\,\,\Psi_3 = 0,\,\,\phi_2 = 0,\,\,
\Phi_{i2} = \overline{\Phi}_{2i} = 0,
\]
\begin{equation}
\label{N2dat}
\Delta\,\phi_1 = 0,\,\,
\Delta\,\phi_0 = \delta\,\phi_1 - 2\,\tau\,\phi_1,\,\,
\phi_0 = u\,(\delta\,\phi_1 - 2\,\tau\,\phi_1),\,\,
\rho = u\,(\overline{\delta}\,\tau - 2\,\alpha\,\tau - \Psi_2)
\quad\mbox{on}\quad {\cal N}'_2.
\end{equation}
\end{lemma}

\noindent{\bf Proof}{\ } In our gauge the relations $\Psi_0 = 0$, 
$\Phi_{00} = k\,\phi_0\,\overline{\phi}_0 = 0$ on ${\cal N}'_1$ are an
immediate consequence of the NP equations and our assumption that $\rho = 0$, 
$\sigma = 0$ on ${\cal N}'_1$. Similarly, the assumptions $\mu = 0$, 
$\lambda = 0$ on ${\cal N}'_2$ imply $\Psi_4 = 0$, 
$\Phi_{22} = k\,\phi_2\,\overline{\phi}_2 = 0$ on ${\cal N}'_2$. The
relation $\Phi_{ij} = k\,\phi_i\,\overline{\phi}_j$ implies the other
statements on the Ricci spinor on ${\cal N}'_1$, ${\cal N}'_2$ and it allows
us to determine $\Phi_{11}$ on ${\cal Z}$ from the data (\ref{Zdata}).

The NP equations involving only the operators $\delta$, $\overline{\delta}$,
the data (\ref{Zdata}), and our gauge conditions allow us to calculate the
functions $\alpha$, $\beta$, $\Psi_1 = 0$, $\Psi_2$, $\Psi_3 = 0$ on 
${\cal Z}$. Then all metric coefficients, spin coefficients, and the Weyl,
Ricci, and Maxwell spinor fields are known on ${\cal Z}$. The remaining
assertions follow by integrating in the appropriate order the NP equations
(cf. also the appendix of \cite{newman:penrose}) involving the operator $D$ 
on ${\cal N}'_1$ and the equations involving the operator $\Delta$ on  
${\cal N}'_2$.  \hfill\fbox{}

\begin{lemma}
\label{Kdat}
A Killing field $K$ as considered in Proposition \ref{Kex} satisfies, up
to a constant factor,
\[
K = r\,z_1 \quad\mbox{on}\quad {\cal N}_1,\,\,\,\,\, 
K = - u\,z_2 \quad\mbox{on}\quad {\cal N}_2.
\]
\end{lemma}

\noindent{\bf Proof}{\ } We note, first, that the statement above is
reasonable, because the vector fields $z_1$, $z_2$ can be defined
globally on ${\cal N}_1$ and ${\cal N}_2$ respectively and the form of
$K$ given above is preserved under rescalings consistent which our
gauge freedom on ${\cal Z}$.

Writing $K = K^i\,z_i$, we have by our assumptions $K = K^1\,z_1$ on 
${\cal N}_1$, $K = K^2\,z_2$ on ${\cal N}_2$ for some smooth functions
$K^1$, $K^2$ which vanish on ${\cal Z}$. To determine their explicit form we
use, in addition to the Killing equation
\begin{equation}
\label{Killing} 
{\cal L}_K\,g_{ij} = \nabla_i\,K_j + \nabla_j\,K_i = 0
\end{equation} 
the identity
\begin{equation}
\label{Kid}
\nabla_i\,\nabla_j\,K_l + K_m\,R^m\,_{ilj} = 
\frac{1}{2}\,\left\{\nabla_i({\cal L}_K\,g_{lj})
+ \nabla_j({\cal L}_K\,g_{li})
- \nabla_l({\cal L}_K\,g_{ij})\right\}
\end{equation}
which holds for arbitrary smooth vector field $K$ and metric $g$ and
which implies, together with eq.(\ref{Killing}), the integrability condition
\begin{equation}
\label{Kint}
\nabla_i\,\nabla_j\,K_l + K_m\,R^m\,_{ilj} = 0.
\end{equation}

The restriction of eq.(\ref{Killing}) to $\zeta$ gives $\nabla_i\,K_j
= 2\,h\,\delta^1\,_{[i}\,\delta^2\,_{j]}$ with $h = z_1\,(K^1) = -
z_2\,(K^2)$. Using this expression to evaluate eq.(\ref{Kint}) on
$\zeta$ for $i = A = 3, 4$, and observing that $\zeta$ is connected we
get $z_A(h) = 0$, whence $h = const.$ on $\zeta$. Since ${\cal Z}$ is
connected the same expression for $\nabla_i\,K_j$ will be obtained
everywhere on ${\cal Z}$ with the same constant $h$. If $h$ were zero,
$K$ would vanish identically by equations (\ref{Killing}) and
(\ref{Kint}).  Since $K$ is assumed to be non-trivial we have $h \neq 0$
and can rescale $K$ to achieve $h = 1$.

Equations (\ref{Killing}), (\ref{Kint}) imply in our gauge
\[
z_1\,(K_2) = - \nabla_2\,K_1,\,\,\,\,\,  
z_1\,(\nabla_2\,K_1) = 0
\quad\mbox{on}\quad {\cal N}'_1,
\]
\[
z_2\,(K_1) = - \nabla_1\,K_2,\,\,\,\,\,  
z_2\,(\nabla_1\,K_2) = 0
\quad\mbox{on}\quad {\cal N}'_2,
\]
which, together with the value of $\nabla_i\,K_j$, $i, j = 1, 2$, 
on ${\cal Z}$ entail our assertion. \hfill\fbox{}

\medskip

Taking into account $\rho = 0$, $\sigma = 0$ on ${\cal N}_1$, $\mu =
0$, $\lambda = 0$ on ${\cal N}_2$, and in particular eq.(\ref{N1dat}), we
immediately get the following.

\begin{lemma}
\label{dKdat}
By calculations which involve only inner derivatives on the respective
null hypersurface one obtains from eq.(\ref{Kdat}) 
\begin{equation}
\label{1nablaKdat}
\nabla_i\,K_j = \delta^1\,_i\,\delta^2\,_j,
\,\,\,\,\,i \neq 2, \quad\mbox{on}\quad {\cal N}_1,
\end{equation}
\begin{equation}
\label{2nablaKdat}
\nabla_i\,K_j = - \delta^2\,_i\,\delta^1\,_j
+ u\,\tau\,\delta^3\,_i\,\delta^1\,_j
+ u\,\overline{\tau}\,\delta^4\,_i\,\delta^1\,_j,\,\,\,\,\,i \neq 1,
\quad\mbox{on}\quad {\cal N}_2.
\end{equation} 
\end{lemma}

Equation (\ref{Kint}) implies the hyperbolic system
\begin{equation}
\label{Kprop}
\nabla_i\,\nabla^i\,K_l - K_m\,R^m\,_l = 0,
\end{equation}
and the initial data for the Killing field we wish to construct are given by
Lemma \ref{Kdat}. Both have an invariant meaning.

\begin{lemma}
\label{Kexist}
There exists a unique smooth solution, $K$, of eq.(\ref{Kprop})
on $D^{+}$ which takes on ${\cal N}_1 \cup {\cal N}_2$ the values given in
Lemma \ref{Kdat}.
\end{lemma}

\noindent{\bf Proof}{\ }
The uniqueness of the solution is an immediate consequence of standard
energy estimates. The results in \cite{mzh} or \cite{rendall} entail the
existence of a unique smooth solution of eq.(\ref{Kprop}) for the data
given in Lemma \ref{Kdat} on an open neighborhood of $\zeta$ in 
$D \cap D^{+}({\cal N}'_1 \cup {\cal N}'_2)$. These local solutions can be
patched together to yield a solution in some neighborhood of ${\cal Z}$ in
$D$. Because of the linearity of eq.(\ref{Kprop}) this solution can be
extended (e.g. by a patching procedure) to all of $D^{+}$. \hfill\fbox{}

\begin{lemma}
\label{K=Killling}
The vector field $K$ of Lemma \ref{Kexist} satisfies
${\cal L}_K\,g = 0$ and ${\cal L}_K\,F = 0$ on $D^{+}$.
\end{lemma}

\noindent{\bf Proof}{\ }
The equations above need to be deduced from the structure of the data in 
Lemma \ref{Kdat} and from eq.(\ref{Kprop}). Applying 
$\nabla_j$ to (\ref{Kprop}) and commuting derivatives we get 
\begin{equation}
\label{propLKg}
\nabla_i\,\nabla^i\,({\cal L}_Kg_{jl}) = 2\,{\cal L}_K\,R_{jl} 
+ 2\,R^i\,_{jl}\,^{k}\,({\cal L}_Kg_{ik})
- 2\,R^i\,_{(j}\,({\cal L}_Kg_{l)i}). 
\end{equation}
The Einstein equations give
\begin{equation}
\label{LKRic}
{\cal L}_K\,R_{ij} = k'\,\left\{2\,F_{(i}\,^l\,({\cal L}_K\,F_{j)l})
- \frac{1}{2}\,g_{ij}\,({\cal L}_K\,F_{kl})\,F^{kl}
- ({\cal L}_Kg_{kl})\,F_i\,^k\,F_j\,^l \right. 
\end{equation}
\[
\left. - \frac{1}{4}\,({\cal L}_Kg_{ij})\,F_{kl} F^{kl}
+ \frac{1}{2}\,g_{ij}\,({\cal L}_K\,g_{kl})\,F^{km}\,F^l\,_m \right\}.
\]
The identity $d\,{\cal L}_K\,F = d\,(i_K\,d\,F + d\,i_K\,F)$
together with Maxwell's equations implies
\begin{equation}
\label{1LKM}
\nabla_{[i}\,({\cal L}_K\,F_{jl]}) = 0.
\end{equation}
Applying ${\cal L}_K$ to the second part of Maxwell's equations
and using the identity (\ref{Kid}) as well as the fact that $K$
solves eq.(\ref{Kprop}), we get 
\begin{equation}
\label{2LKM} 
\nabla^i\,({\cal L}_K\,F_{ik}) = F^{jl}\,\nabla_j({\cal L}_K\,g_{lk})
+ ({\cal L}_K\,g_{jl})\,\nabla^j\,F^l\,_k.
\end{equation}

Substituting eq.(\ref{LKRic}) in eq.(\ref{propLKg}), we can view the system
(\ref{propLKg}), (\ref{1LKM}), (\ref{2LKM}) as a homogeneous linear system for the
unknowns ${\cal L}_K\,g$, ${\cal L}_K\,F$. This system implies a
linear symmetric hyperbolic system for the unknowns ${\cal L}_K\,g$,
$\nabla\,{\cal L}_K\,g$, ${\cal L}_K\,F$ (cf. \cite{friedrich}). 

We shall show now that these unknowns vanish on 
${\cal N}_1 \cup {\cal N}_2$.  The standard energy estimates for symmetric
hyperbolic systems then imply that the fields vanish in fact on $D^{+}$,
which will prove our lemma and thus Proposition \ref{Kex}. Equation
(\ref{Kprop}) restricted to ${\cal N}'_1$ reads
\[
0 = \nabla_i\,\nabla^i\,K_l - K_m\,R^m\,_l
= 2\,(\nabla_1\,\nabla_2\,K_l - \nabla_3\,\nabla_4\,K_l)
- K_m\,(R^m\,_l + R^m\,_{l21} - R^m\,_{l43}).
\]
Using this equation together with eqs.(\ref{N1dat}) and (\ref{1nablaKdat}) and our gauge
conditions, we obtain by a direct calculation a system of ODE's of the form
\[
\nabla_1\,(\nabla_{(i}\,K_{j)}) = H_{ij}(\nabla_{(k}\,K_{l)}), 
\]
on the null generators of ${\cal N}'_1$. Here $H_{ij}$ is a linear function
of the indicated argument (suppressing the dependence on the points of
${\cal N}'_1$). Since $\nabla_{(k}\,K_{l)} = 0$ on ${\cal Z}$, we  conclude
that ${\cal L}_K\,g = 0$ on ${\cal N}_1$. An analogous argument involving
eqs.(\ref{N2dat}) and (\ref{2nablaKdat}) shows that ${\cal L}_K\,g = 0$ on
${\cal N}_2$. It follows in particular that
$\nabla\,{\cal L}_K\,g = 0$ on ${\cal Z}$.

Writing $({\cal L}_K\,F)_{AA'BB'} = \epsilon_{A'B'}\,p_{AB}
+ \epsilon_{AB}\,\overline{p}_{A'B'}$
and using (\ref{Kdat}) we find on ${\cal N}'_1$ in NP notation
\[
p_0 = r\,D\,\phi_0 + \phi_0,\,\,\,\,\,
p_1 = r\,D\,\phi_1,\,\,\,\,\,
p_2 = r\,D\,\phi_2 - \phi_2.
\]      
It follows from eq.(\ref{N1dat}) that $p_{AB}$, whence ${\cal L}_K\,F$, vanishes on
${\cal N}_1$. An analogous argument involving eq.(\ref{N2dat}) shows that 
${\cal L}_K\,F$ vanishes on ${\cal N}_2$.

Observing in eqs.(\ref{LKRic}) and (\ref{propLKg}) that ${\cal L}_K\,F = 0$, 
${\cal L}_K\,g = 0$, whence also $\nabla_l(\nabla_{(i}\,K_{j)}) = 0$ for 
$l \neq 2$ on ${\cal N}'_1$, we obtain there 
\[
0 = \nabla_k\,\nabla^k\,(\nabla_{(i}\,K_{j)})
= 2\,\nabla_1\,(\nabla_2\,(\nabla_{(i}\,K_{j)})). 
\]
We conclude that $\nabla\,{\cal L}_K\,g$, which vanishes on ${\cal Z}$,
vanishes on ${\cal N}_1$. In a similar way it follows that 
$\nabla\,{\cal L}_K\,g = 0$ on ${\cal N}_2$. This completes the proof. 
\hfill\fbox{}


\begin{thebibliography}{99}

\bibitem{hawk1}  S.W. Hawking: \textit{Black holes in general relativity},
Commun. Math. Phys. \textbf{25}, 152-166 (1972)

\bibitem{HE}  S.W. Hawking and G.F.R. Ellis:\textsl{\ The large scale
structure of space-time}, Cambridge University Press (1973)

\bibitem{chrus22}  P.T. Chru\'sciel: \textit{On rigidity of analytic black 
holes},  Commun. Math. Phys. \textbf{189}, 1-7 (1997)

\bibitem{chrus3}  P.T. Chru\'sciel: \textit{Uniqueness of stationary, 
electro-vacuum black holes revisited}, Helv. Phys. Acta \textbf{69}, 529-552
(1996)

\bibitem{israel1}  W. Israel: \textit{Event horizons in static vacuum
space-times}, Phys. Rev. \textbf{164}, 1776-1779 (1967)

\bibitem{israel2}  W. Israel: \textit{Event horizons in static electrovac
space-times}, Commun. Math. Phys. \textbf{8}, 245-260 (1968)

\bibitem{carter1}  B. Carter: \textit{Axisymmetric black hole has only two
degrees of freedom}, Phys. Rev. Lett. \textbf{26}, 331-333 (1971)

\bibitem{carter2}  B. Carter: \textit{Black hole equilibrium states, in:
Black Holes}, C. de Witt and B. de Witt (eds.) Gordon and Breach, New York,
London, Paris (1973)

\bibitem{mazur}  P.O. Mazur: \textit{Proof of uniqueness of the Kerr-Newman
black hole solutions}, J. Phys. A: Math. Gen. \textbf{15}, 3173-3180 (1982)

\bibitem{bunting}  G.L. Bunting: \textit{Proof of the uniqueness conjecture
for black holes}, Ph. D. Thesis, University of New England, Admirale (1987)

\bibitem{im1} V. Moncrief and J. Isenberg: \textit{Symmetries of
cosmological Cauchy horizons}, Commun. Math. Phys. \textbf{89}, 387-413
(1983)

\bibitem{im2}  J. Isenberg and V. Moncrief: \textit{Symmetries of
cosmological Cauchy horizons with exceptional orbits}, J. Math. Phys. 
\textbf{26}, 1024-1027 (1985)

\bibitem{P2}  R. Penrose:\textit{\ Singularities an time asymmetry} in 
\textsl{General relativity; An Einstein centenary survey}, eds. S.W. 
Hawking, W. Israel, Cambridge University Press (1979)

\bibitem{racz} I. R\'acz: {\it On further generalization of the
rigidity theorem for spacetimes with a stationary event horizon or a
compact Cauchy horizon}, in preparation.

\bibitem{rw1}  I. R\'{a}cz and R.M. Wald: \textit{Extension of spacetimes
with Killing horizon}, Class. Quant. Grav. \textbf{9}, 2643-2656 (1992)

\bibitem{rw2}  I. R\'{a}cz and R.M. Wald: \textit{Global extensions of
spacetimes describing asymptotic final states of black holes, }Class. Quant.
Grav. \textbf{13}, 539-553 (1996)

\bibitem{wald} R.M. Wald: \textsl{\ General relativity},
University of Chicago Press, Chicago (1984)

\bibitem{CW2}  P.T. Chru\'sciel and R.M. Wald: 
\textit{Maximal hypersurfaces in asymptotically flat spacetimes}, 
Commun. Math. Phys. \textbf{163}, 561-604 (1994)

\bibitem{FSW}  J.L. Friedman, K. Schleich and D.M. Witt: 
\textit{Topological censorship}, Phys. Rev. Lett. \textbf{71}, 1486-1489 (1993)

\bibitem{CW}  P.T. Chru\'sciel and R.M. Wald: 
\textit{On the topology of stationary black holes}, 
Class. Quant. Grav. \textbf{11}, L147-L152 (1994)

\bibitem{G1}  G.J. Galloway: \textit{On the topology of the domain of outer 
communication}, Class. Quant. Grav. \textbf{12}, L99-L101 (1995)

\bibitem{G2}  G.J. Galloway: \textit{A ``finite infinity'' version of the 
FSW topological censorship}, Class. Quant. Grav. \textbf{13}, 1471-1478 (1996)

\bibitem{geroch}  R. Geroch:
\textit{A method for constructing solutions of Einstein's equations},
J. Math. Phys. \textbf{12}, 918-924 (1971)

\bibitem{W2}  R.M. Wald: \textsl{\ Quantum field theory on curved spacetimes},
University of Chicago Press, Chicago (1994)

\bibitem{mill}  J.G. Miller: \textit{Global analysis of the Kerr-Taub-NUT
metric}, J. Math. Phys. \textbf{14}, 486-494 (1973)

\bibitem{monc1}  V. Moncrief: \textit{Infinite-dimensional family of vacuum
cosmological models with Taub-NUT (}Newman-Unti-Tamburino\textit{)-type
extensions}, Phys. Rev. D. \textbf{23}, 312-315 (1981)

\bibitem{monc2}  V. Moncrief: \textit{Neighborhoods of Cauchy horizons in
cosmological spacetimes with one Killing field}, Ann. of Phys. \textbf{141}.
83-103 (1982)

\bibitem{se}  H. Seifert: \textit{Topologie dreidimensionaler gefaserter
R\"{a}ume}, Acta. Math. \textbf{60}, 147-238 (1933). English translation 
in: H. Seifert, W. Threlfall: \textit{A Textbook of Topology}, Academic Press,
New York, 1980. 

\bibitem{ep}  D.B.A. Epstein: \textit{Periodic flows on three-manifolds},
Ann. Math. \textbf{95}, 66-81 (1972)

\bibitem{newman:penrose} E. Newman, R. Penrose:\textit{An Approach to 
Gravitational Radiation by a Method of Spin coefficients.}, 
J. Math. Phys. \textbf{3} 566 - 578 (1962), \textbf{4}, 998 (1963)  

\bibitem{mzh} H. M\"uller zum Hagen:\textit{Characteristic initial value
problem for hyperbolic systems of second order differential systems},
Ann. Inst. Henri Poincar\'e  \textbf{53}, 159 - 216 (1990)

\bibitem{rendall}  A.D. Rendall:\textit{Reduction of the characteristic
initial value problem to the Cauchy problem and its applications to the
Einstein equations}, Proc. R. Soc. Lond. A \textbf{427}, 221-239 (1990)

\bibitem{friedrich}  H. Friedrich:\textit{On the Global Existence and the
Asymptotic Behaviour of Solutions to the Einstein-Maxwell-Yang-Mills
Equations}, J. Diff. Geom. \textbf{34}, 275 - 345 (1991)


\end{thebibliography}
\end{document}